\documentclass[sigconf,screen]{acmart}

\copyrightyear{2025}
\acmYear{2025}
\setcopyright{acmlicensed}\acmConference[FSE Companion '25]{33rd ACM International Conference on the Foundations of Software Engineering}{June 23--28, 2025}{Trondheim, Norway}
\acmBooktitle{33rd ACM International Conference on the Foundations of Software Engineering (FSE Companion '25), June 23--28, 2025, Trondheim, Norway}
\acmDOI{10.1145/3696630.3728584}
\acmISBN{979-8-4007-1276-0/2025/06}

\usepackage{balance}
\usepackage{hyperref}
\usepackage{cleveref}
\usepackage{soul}
\usepackage{url}
\usepackage{listings}
\lstdefinelanguage{Kotlin}{
  comment=[l]{//},
  commentstyle={\color{gray}\ttfamily},
  emph={filter, first, firstOrNull, forEach, lazy, map, mapNotNull, println},
  emphstyle={\color{orange!60!black}},
  identifierstyle=\color{black},
  keywords={!in, !is, abstract, actual, annotation, as, as?, break, by, catch, class, companion, const, constructor, continue, crossinline, data, delegate, do, dynamic, else, enum, expect, external, false, final, finally, for, fun, get, if, import, in, infix, init, inline, inner, interface, internal, is, lateinit, noinline, null, object, open, operator, out, override, package, param, private, property, protected, public, reified, return, return@, sealed, set, super, suspend, tailrec, this, throw, true, try, typealias, typeof, val, var, vararg, when, where, while},
  keywordstyle={\color{blue!60!black}\bfseries},
  morecomment=[s]{/*}{*/},
  morestring=[b]",
  morestring=[s]{"""*}{*"""},
  ndkeywords={@Deprecated, @JvmField, @JvmName, @JvmOverloads, @JvmStatic, @JvmSynthetic, Array, Byte, Double, Float, Int, Integer, Iterable, Long, Runnable, Short, String, Any, Unit, Nothing},
  ndkeywordstyle={\color{BurntOrange}\bfseries},
  sensitive=true,
  basicstyle={\ttfamily \color{green!60!black} \small}
}
\usepackage{multirow}
\newcommand{\code}[1]{\texttt{#1}}
\usepackage{xspace}
\newcommand{\jcstress}{\textit{jcstress}\xspace}
\newcommand{\herdlitmus}{\textit{herdtools/litmus7}\xspace}
\newcommand{\litmuskt}{\textit{LitmusKt}\xspace}

\newcommand{\eg}{\emph{e.g.,}\xspace}

\title{LitmusKt: Concurrency Stress Testing for Kotlin}

\author{Denis Lochmelis}
\affiliation{
  \institution{\textit{Constructor University Bremen}}
  \institution{\textit{JetBrains Research}}
  \city{Bremen}
  \country{Germany}
}
\email{dlochmelis@constructor.university}

\author{Evgenii Moiseenko}
\affiliation{
  \institution{\textit{JetBrains Research}}
  \city{Belgrade}
  \country{Serbia}
}
\email{evgeniy.moiseenko@jetbrains.com}

\author{Yaroslav Golubev}
\affiliation{
  \institution{\textit{JetBrains Research}}
  \city{Belgrade}
  \country{Serbia}
}
\email{yaroslav.golubev@jetbrains.com}

\author{Anton Podkopaev}
\affiliation{
\institution{\textit{Constructor University Bremen}}
  \city{Bremen}
  \country{Germany}
}
\affiliation{
  \institution{\textit{JetBrains Research}}
  \city{Amsterdam}
  \country{The Netherlands}
}
\email{apodkopaev@constructor.university}

\begin{abstract}

We present \litmuskt---the first tool for litmus testing concurrent programs in Kotlin. The tool's novelty also lies in the fact that Kotlin is a \textit{multiplatform} language, \textit{i.e.}, it compiles into multiple platforms, which means that the concurrency has to be tested on several of them. Our tool allows writing litmus tests in a single custom DSL, and these tests are then run in Kotlin/Native and Kotlin/JVM, two main platforms for concurrent programming in Kotlin. Using \litmuskt, we discovered novel bugs in the Kotlin compiler, which we then fixed and they are no longer present. Moreover, \litmuskt was integrated into the CI pipeline for Kotlin. \litmuskt is available on GitHub: \url{https://github.com/JetBrains-Research/litmuskt}. The demo is available on YouTube: \url{https://youtu.be/oWCZp_Huwss}.

\end{abstract}

\begin{document}



\maketitle

\begin{figure}
\vspace{0.5cm}
    \centering
    \ttfamily
    \begin{tabular}{c||c}
        \multicolumn{2}{c}{x, y, r1, r2 = 0} \\[+0.5em]
        x = 1 & y = 1 \\
        r1 = y & r2 = x \\[+0.5em]
        \multicolumn{2}{c}{(r1,r2) $\stackrel{?}{=}$ (0,0) }
    \end{tabular}
    \caption{\textit{Store Buffering} litmus test, a sample program that demonstrates weak behavior.}
    \vspace{-0.4cm}
    \label{fig:sb-schematic}
\end{figure}

\section{Introduction}
\label{sec:intro}
Concurrent programming is notoriously hard and error-prone~\cite{rajsbaum2018mastering}.
In large part, this is due to modern compilers and hardware carrying out various optimizations on programs, such as reordering of instructions and out-of-order execution.
Due to these processes, concurrent programs can exhibit counterintuitive behaviors that cannot be observed in the case of single-threaded programs.

As a classic example, consider the \textit{Store Buffering} program in~\Cref{fig:sb-schematic}, where the program first initializes locations $x$, $y$, $r1$, and $r2$ to $0$ and then spawns two threads. Suppose we try to run this program on a single-core machine without compiler optimizations.
This would result in some interleaving of read and write operations,
as defined by the model of sequential consistency~\cite{Lamport:TC79}.
Notice that both threads start with a write operation either to location $x$ or to location $y$, implying that $r1$ and $r2$ both being equal to $0$ at the end of execution is seemingly impossible.
However, such behavior can be observed in a real-world multi-core setting due to compilers reordering instructions in threads (\eg GCC may do so~\cite{Vafeiadis-al:POPL15}), and such outcomes are referred to as \emph{weak}~\cite{Alglave-al:TOPLAS14}.

The set of allowed outcomes for a concurrent program in a language is usually specified by the language's \emph{memory model}~\cite{Batty-al:POPL11,Manson-al:POPL05}.
Developers use the model to reason about their concurrent programs, and compiler creators use it as a specification to which they must adhere.
Unfortunately, the latter is not always easy, and there are bugs in implementations of concurrency primitives, \eg in a popular implementation of JVM~\cite{Liu-al:ECOOP22}.
Such bugs in compilers and virtual machines may have a much bigger impact than in user-level applications~\cite{Oberhauser-al:NETYS21,Liu-al:ECOOP22}, therefore, they require thorough testing.

In this work, we focus on testing the implementation of concurrency in Kotlin~\cite{kotlin}, a popular language developed by JetBrains.
The fact that Kotlin programs can be compiled to different platforms---JVM, Native via LLVM, JavaScript, and WebAssembly---makes it a particularly interesting target for concurrency testing, since each of these platforms has a different underlying memory model. This in turn requires testing that the behavior of the same Kotlin program on different platforms is uniform.
At the moment, Kotlin lacks a formally defined memory model, so there is no solid reference to test the implementation against. However, there are some well-known general patterns that can still be tested and that we decided to discuss with Kotlin's language designers and compiler engineers.

There are different ways to test the compiler's memory model.
A popular method is to use the \textit{litmus testing} technique~\cite{Alglave-al:TOPLAS14, jcstress}.
It works as follows.
First, we write specially crafted small concurrent programs, called \textit{litmus tests}, which are designed to highlight specific behaviors or potential bugs.
An example of such program, the \textit{Store Buffering} litmus test, is shown in~\Cref{fig:sb-schematic}.
Then, we run them repeatedly for some time and record all the results that were produced by these programs.
Finally, we check whether these results are allowed under our assumptions about the language memory model.
If we see a forbidden result, there must be a bug with the language implementation---that is, in the compiler.

In reality, the proper litmus testing process is a bit more involved. Crucially, it is not enough to simply rerun the test multiple times, because as mentioned previously, concurrent programs can exhibit weak behaviors that require certain thread interactions and are considerably rare. Therefore, to maximize the chances of observing all possible behaviors, certain techniques should be applied, and to do that, a litmus testing tool should be employed.
There exist several well-known tools like that. One is \jcstress~\cite{jcstress}, which takes litmus tests written in Java and runs them in order to test the conformance of the JVM to the Java Memory Model (JMM). Another tool is \herdlitmus~\cite{Alglave-al:TOPLAS14}, which takes litmus tests written in Assembly and checks the memory model of the processor which is running the test. Unfortunately, neither can be reused fully for our case with Kotlin. One key feature of Kotlin is it being \textit{multiplatform}, which means that we need to perform litmus testing with several platforms, and for Kotlin/Native, no existing tool can help us.

To overcome the existing gap in research and practice, we built \litmuskt, the first tool designed specifically for litmus testing the Kotlin compiler. Specifically, our tool has the following features.

\begin{itemize}
    \item Firstly, \litmuskt runs litmus tests written in Kotlin and is able to support different platforms (in particular, Kotlin/Native and Kotlin/JVM, as these two offer the most comprehensive support of concurrent programming). If the tool finds a discrepancy between the behaviors on different platforms, this could mean a potential bug in the Kotlin compiler.
    \item Secondly, \litmuskt employs the methods used by other known litmus tools to increase the frequency of rare concurrent behaviors, including the weak ones. Even without comparing different platforms, finding all unusual behaviors on their own proved to be valuable.
    \item Finally, \litmuskt provides a way for declaring new litmus tests in the form of a DSL. The syntax balances simplicity and flexibility for writing arbitrary tests.
\end{itemize}

Using this tool, we were able to find new bugs in the Kotlin/Native compiler related to unsafe publication of objects, which we discuss in detail in Section~\ref{findings}. Our fix for these bugs was merged into Kotlin~\cite{fix-commit}, and they are no longer present. In addition, our work has been acknowledged by the Kotlin/Native development team, and we integrated \litmuskt into the Kotlin CI testing pipeline~\cite{ci-commit}. We believe that our tool is of value to both researchers and practitioners for continuing the study of concurrency in Kotlin, but also broader---for litmus testing in several platforms at the same time and finding novel types of \textit{multiplatform} bugs.

\litmuskt is openly available on GitHub: \url{https://github.com/JetBrains-Research/litmuskt}. The demonstration video is available on YouTube: \url{https://youtu.be/oWCZp_Huwss}.
\section{Implementation}
\label{sec:implementation}

To highlight how the tool operates, we first demonstrate an example of a \litmuskt test and then discuss the implementation details of the framework.

\subsection{Example of a \litmuskt Test}
\label{subsec:test-example}

In Listing~\ref{lst:sb_dsl}, one can see an example of a litmus test written in the \litmuskt DSL syntax (the same \textit{Store Buffering} test presented in Figure~\ref{fig:sb-schematic}).
The test is declared with the \code{litmusTest} function and consists of several parts:
the definition of a shared state, 
running code for several threads,
and the specification of test outcomes.

\begin{lstlisting}[
    caption={\textit{Store Buffering} test written in \litmuskt syntax.\vspace{-0.4cm}},
    captionpos=b,
    label={lst:sb_dsl},
    float,
    language=Kotlin
]
val StoreBuffering = litmusTest({
    object : LitmusIIOutcome() { var x, y = 0, 0 }
}) {
    thread { x = 1; r1 = y }
    thread { y = 1; r2 = x }
    spec {
        accept(0, 1)
        accept(1, 0)
        accept(1, 1)
        interesting(0, 0)
    }
}
\end{lstlisting}

The first lambda argument defines the object to be used as a shared state between the threads.
This object declares two shared variables: \code{x} and \code{y}.
This particular state object also inherits from the predefined class \code{LitmusIIOutcome}\footnote{
  \code{II} stands for \code{IntInt}, the naming scheme is taken from~\jcstress.
}
that brings into the scope two additional integer variables \texttt{r1} and \texttt{r2}.\footnote{
  \code{rN} stands for \emph{register \#N}---by convention in the weak memory literature,
  registers denote \emph{local variables of threads},
  as opposed to \emph{variables shared between several threads}.
}
Their final state is automatically used as the test outcome.
In general, it is possible to either use different \code{LitmusXXXOutcome}-s, which are optimized for convenience and performance, or declare a completely custom outcome using the \code{outcome} section of the DSL.

Next, the \code{thread} DSL sections simply define the running code for each thread.
Finally, following the convention used in \jcstress,
the \code{spec} DSL section specifies what outcomes are
\begin{itemize}
\item \textcolor{gray}{acceptable}---sequentially consistent outcomes, \emph{i.e.,} reproducible as an interleaving of threads;
\item \textcolor{orange}{interesting}---weak but tolerable outcomes according to the expected
semantics of the programming language;
\item or \textcolor{red}{forbidden}---weak intolerable outcomes, indicating bugs.
\end{itemize}
For example, as mentioned in Section~\ref{sec:intro}, \texttt{(0,0)} is a tolerable result in \textit{Store Buffering}, and so it is marked as \textcolor{orange}{interesting} in Listing~\ref{lst:sb_dsl}. 

When the test is written, the user can run it using the provided command line interface (CLI).
The CLI allows setting various parameters for the run,
such as testing duration, number of parallel runs, thread CPU affinity, and so on. All parameters available to the CLI are described in the project's README~\cite{readme}.

Once the test finishes, \litmuskt produces a resulting report
(see an example in~\Cref{fig:sb_result}).
The resulting table has a separate row for each encountered outcome:
the columns of the table describe the category of an outcome,
the number of times it was encountered, and its frequency. The overall status of the entire test is \textcolor{red}{forbidden} if at least one \textcolor{red}{forbidden} outcome was found, and otherwise it is \textcolor{orange}{interesting} if at least one \textcolor{orange}{interesting} outcome was found.

The produced output is meant for manual analysis. For example, as mentioned in Section~\ref{sec:intro}, the \textit{Store Buffering} test can easily produce a weak outcome, and this fact is clearly indicated in the sample output. It can also be seen that the weak outcome is not necessarily the rarest. Let us now describe how \litmuskt operates.

\begin{table}[t]
\centering

\begin{tabular}{c|c|c|c}
\toprule
\textbf{Outcome}  & \textbf{Type}  & \textbf{Count} & \textbf{Frequency}    \\ \midrule
 (1, 0)  & \textcolor{gray}{ACCEPTABLE}   & 505,376 &  50.537\% \\
 (0, 1)  &  \textcolor{gray}{ACCEPTABLE}   & 493,675 &  49.367\% \\
 (0, 0)  & \textcolor{orange}{INTERESTING} &  942   &  0.0942\% \\
 (1, 1)  &  \textcolor{gray}{ACCEPTABLE}   &   7    &  <0.001\% \\ \midrule
 \multicolumn{4}{c}{total count: 1,000,000, overall status: \textcolor{orange}{INTERESTING}} \\
 \bottomrule
\end{tabular}
\vspace{0.2cm}
\caption{Sample \litmuskt output after running the \textit{Store Buffering} test.}
\vspace{-0.8cm}
\label{fig:sb_result}

\end{table}

\subsection{Kotlin/Native Test Runner}
\label{subsec:runner}

For Kotlin/Native, no solutions exist, and so we had to develop our own test runner from scratch. To increase the probability of triggering weak behaviors,
\litmuskt employs a set of standard techniques used by other prominent litmus testing tools,
such as \jcstress~\cite{jcstress} and \herdlitmus~\cite{Alglave-al:TOPLAS14}.

\begin{figure}[t]
\centering
\includegraphics[width=3.4in]{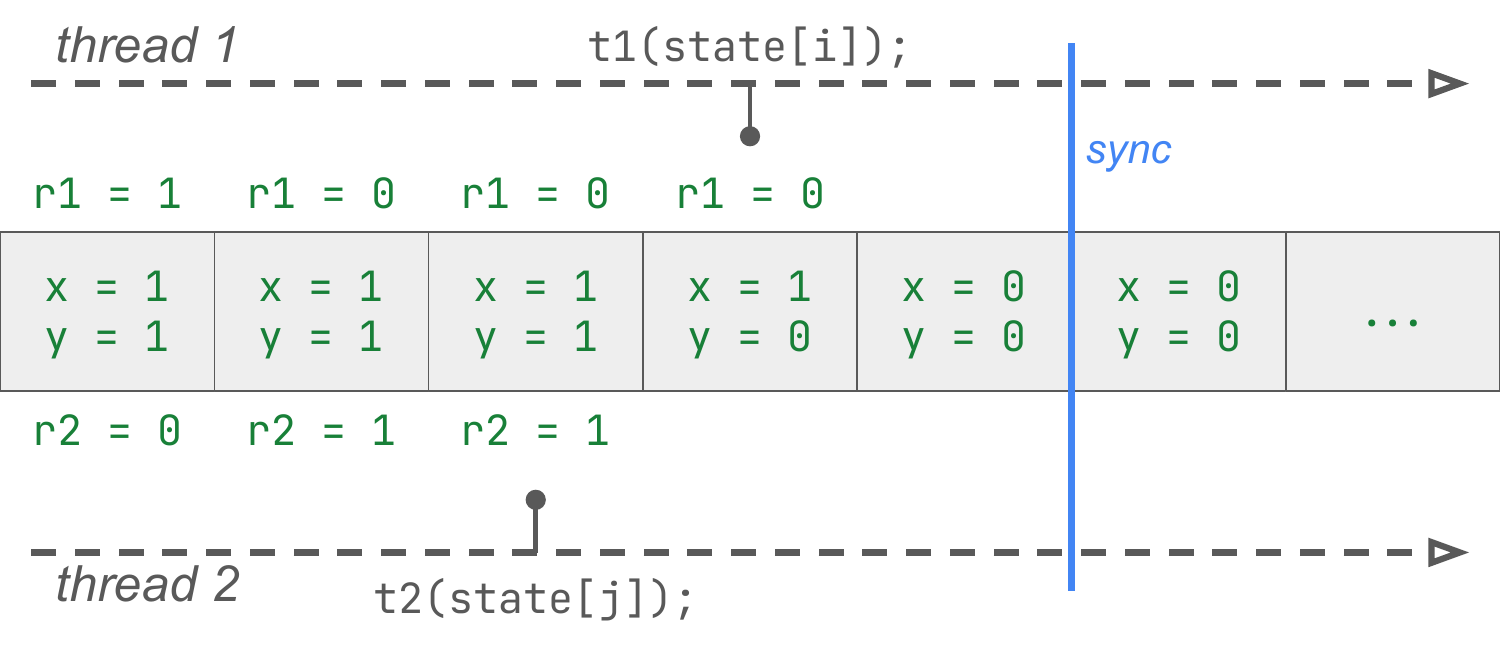}
\caption{A schematic of the \litmuskt test running process, using \emph{Store Buffering} as an example. One gray rectangle represents a single test state. We allocate an array of them. Then, test threads sequentially execute their functions on this array of states. The threads also regularly synchronize on barriers. }

\label{fig:core_schematic}
\end{figure}

The general principle of \litmuskt test runner is illustrated by~\Cref{fig:core_schematic}, using the same \textit{Store Buffering} test as an example.
First, the runner initializes a large array of test states.
Then, it launches several threads, as declared in the litmus test being run. These threads iterate over the array of states and simultaneously execute their code on each state instance.
To prevent one thread from significantly running ahead of others, every \code{N} iterations, threads perform a wait on a barrier to synchronize with each other.
Finally, after the threads have finished iterating, the outcomes are collected from the states and quickly analyzed to produce the overall test result.

With this setup, the threads modify different states simultaneously many times over, significantly increasing the likelihood that some concurrent interaction occurs. This general principle is further enhanced by other techniques, \eg it is possible to directly set CPU affinity for each thread to ensure complex interactions.

\subsection{Kotlin/JVM Test Runner Based on \jcstress}
\label{subsec:jcstress-interop}

The ability of running tests on both Kotlin/Native and Kotlin/JVM and comparing their results was one of the main requirements of \litmuskt.
At the same time, for Kotlin/JVM compiler, reusing the established tool \jcstress~\cite{jcstress} would be advantageous, as it already implements many advanced techniques and has a proven track record of uncovering concurrency bugs in various JVM implementations~\cite{jcstress_workshop}.
Therefore, we decided to support \jcstress in the \litmuskt as a backend for running tests on Kotlin/JVM. 

However, such integration turned out to be technically challenging.
The primary issue was that \jcstress heavily relies on using Java Reflection,
a feature not fully supported in Kotlin Multiplatform at the moment.
This limitation made it difficult to adapt litmus tests written in our Kotlin syntax for direct use with \jcstress.
Neither was it easy to embed \jcstress as a library into the \litmuskt pipeline, and,
specifically, to trigger internal \jcstress code generation passes.

Because of this, we decided to treat \jcstress as a ``black box'' for running tests.
The resulting pipeline operates as follows.
First, we compile the Kotlin litmus tests into JVM bytecode and pack them into a JAR archive.
Second, we use a simple code generator to create a \jcstress-compliant
Java wrapper class for each litmus test.
The generated code delegates the execution to the Kotlin-compiled litmus test code
from the JAR library created on the first step.
Next, \jcstress runs on these generated Java tests and produces the reports in the HTML format.
Finally, we parse these reports into the \litmuskt internal format.
This entire process is encapsulated within a common test runner interface,
enabling seamless integration of the \jcstress-based test runner in \litmuskt.

\paragraph{\bf Summary.}
As demonstrated, \litmuskt conforms to all the requirements that we set forth. It allows writing litmus tests in a flexible yet concise syntax, runs them using efficient techniques, and supports multiple platforms by delegating to \jcstress on JVM and using a custom runner for Native, unifying them under a common interface. 

\section{Analysis} \label{findings}

\begin{table}[t]

\newcommand{\accepted}{\textcolor{gray}{ACC}\xspace}
\newcommand{\interesting}{\textcolor{orange}{INT}\xspace}
\newcommand{\forbidden}{\textcolor{red}{FORB}\xspace}
\centering

\centering

\begin{tabular}{ccccc}
\toprule
\multirow{2}{*}{\textbf{Litmus test}} & \multicolumn{2}{c}{\textbf{Native}} & \multicolumn{2}{c}{\textbf{JVM}} \\ \cmidrule(lr){2-3} \cmidrule{4-5}
            & \textbf{x86}          & \textbf{Arm}           & \textbf{x86}           & \textbf{Arm} \\ 
\midrule
ATOM    & \accepted     & \accepted     & \accepted     & \accepted \\ 
SB      & \interesting  & \interesting  & \interesting  & \interesting\\
SB volatile & \accepted & \accepted     & \accepted     & \accepted \\
MP      & \interesting  & \accepted     & \interesting  & \interesting \\
MP volatile & \accepted & \accepted     & \accepted     & \accepted \\
MP DRF  & \accepted     & \accepted     & \accepted     & \accepted \\
CoRR    & \accepted     & \accepted     & \accepted     & \accepted \\
CoRR CSE & \accepted    & \accepted     & \accepted     & \accepted \\
IRIW    & \accepted     & \accepted     & \interesting  & \interesting \\
IRIW volatile & \accepted & \accepted   & \accepted     & \accepted \\
UPUB    & \accepted     & \forbidden    & \accepted     & \accepted \\
OOTA    & \accepted     & \accepted     & \accepted     & \accepted \\
LB      & \accepted     & \accepted     & \accepted     & \accepted \\
LB Deps & \accepted     & \accepted     & \interesting  & \interesting\\
\bottomrule
\end{tabular}
\vspace{0.2cm}
\caption{The results of our analysis. \accepted indicates that the results are \textcolor{gray}{acceptable}, \forbidden---that at least one result is \textcolor{red}{forbidden}, \interesting---that at least one result is \textcolor{orange}{interesting}.}
\vspace{-0.3cm}
\label{fig:results-table}

\end{table}

We ran a number of relatively standard litmus tests~\cite{Moiseenko-al:Programming2021} with \litmuskt on two machines: \textit{x86} machine (Intel i7-12700K CPU) and \textit{Arm} machine (Apple M1 CPU). The results are presented in~\Cref{fig:results-table}. Many of the tests that theoretically could show some weak behavior did not do so in practice. This is expected, because Kotlin memory model is not formalized, and as such, it cannot be determined in advance which tests may show weak behaviors in practice. However, some tests still showed noteworthy behaviors, and we will now discuss those tests in more detail.

Firstly, the \textit{Unsafe Publication} test (\textit{UPUB} for short) demonstrates \textcolor{red}{forbidden} outcomes. Its schematic description is shown in~\Cref{fig:upub-schematic}. Here, one thread initializes a class instance, while another thread simultaneously tries to read its field.
When we run this test, we expect to see either no instance in the reading thread (\textit{i.e.}, \texttt{null}), or a fully initialized instance, or a partially initialized class with default values for fields. However, as shown in~\Cref{fig:upub-results}, when running with Kotlin/Native 1.9.0 on an Arm processor, it is possible to see "garbage" values. The problem is exacerbated if we substitute the \texttt{Int} field with a class reference, because when we try to read a reference with a "garbage" address, we get a segmentation fault.

This behavior is clearly not acceptable for a memory-safe language like Kotlin, or in other words, this was a newly discovered bug in the compiler. We investigated it further and then implemented a fix, which was since merged~\cite{fix-commit}. As of Kotlin 1.9.20 and higher, this bug no longer occurs.

\begin{figure}[t]
    \centering
    \ttfamily
    \begin{tabular}{c||c}
        \multicolumn{2}{c}{class IntHolder(var x: Int = 0)} \\
        \multicolumn{2}{c}{h: IntHolder? = null} \\
        \rule{0pt}{1.5em} h = IntHolder() & r1 = h?.x \rule[-1em]{0pt}{0pt} \\
        \multicolumn{2}{c}{r1 $\in$ \{0, null\}  }
    \end{tabular}
        \vspace{-0.2cm}
    \caption{The \textit{Unsafe Publication (UPUB)} test.}
    \vspace{-0.2cm}
    \label{fig:upub-schematic}
\end{figure}

We found more problems regarding \textit{UPUB}, caused by the fact that classes can be published with fields equal to default values. For example, Kotlin has nullable and non-nullable types, and under normal conditions, it should not be possible to read \texttt{null} from a non-nullably-typed variable. However, a similar \textit{UPUB with reference} test can be used to observe the opposite, and we do in fact see it. This fact presents a new challenge for language design.

\begin{table}[t]
\centering

\begin{tabular}{c|c|c|c}
\toprule
\textbf{Outcome}  & \textbf{Type}  & \textbf{Count} & \textbf{Frequency}    \\ \midrule
    null     & \textcolor{gray}{ACCEPTABLE}  & 26,767,684 &  65.607\%\\
      0      & \textcolor{gray}{ACCEPTABLE}  & 14,032,239 & 34.392\%\\
  71817408   & \textcolor{red}{FORBIDDEN} &    12    &  <0.001\%\\
 -1493360416 & \textcolor{red}{FORBIDDEN} &    1     &  <0.001\%\\
 -1560102576 & \textcolor{red}{FORBIDDEN} &    1     &  <0.001\%\\
 \multicolumn{4}{c}{...} \\ \midrule
 \multicolumn{4}{c}{total count: 40,800,000, overall status: \textcolor{red}{FORBIDDEN}} \\
 \bottomrule
\end{tabular}
\vspace{0.2cm}
\caption{Results of the \textit{UPUB} test, Kotlin/Native 1.9.0, Arm processor.}
\vspace{-0.7cm}
\label{fig:upub-results}

\end{table}

Another similar problem arises from using arrays. In this variation of \textit{UPUB}, one thread initializes an array and another tries to read its first element. In Kotlin, \texttt{Array} stores objects, and these objects are also susceptible to the default initialization problem. For example, \texttt{Array<Int>} will store boxed \texttt{Int} objects, and even when the reader thread is able to read the array element, it can still read a garbage value from them, and again crash with a segmentation fault, which we have confirmed.

To summarize, by running \litmuskt, we have discovered several new problems with Kotlin/Native. They have piqued the interest of the Kotlin/Native development team, and it was decided to embed \litmuskt as a part of the Kotlin CI testing pipeline~\cite{ci-commit}.
\section{Future work}
\label{sec:future_work}

During our experiments with \litmuskt, we made a couple of observations that suggest some meaningful directions for future work.

One potential path for research is highlighted by the \textcolor{orange}{interesting} outcomes of the \textit{Message Passing (MP)} litmus test. This test looks similar to \textit{Store Buffering}, even though it tests a different memory model aspect. What is so peculiar about its results in Table~\ref{fig:results-table} is that we observe the weak outcome for Native only on x86 processors, which are generally much more restrictive in terms of weak behaviors compared to Arm. This observation can only be reasonably explained by aggressive compiler optimizations happening due to obscure reasons that cannot be directly controlled. Based on this idea, we believe that fuzzing litmus tests via code mutation to trigger random compiler optimizations has a decent chance of yielding more weak behaviors, and therefore finding more bugs.

Another scope for future work is adding further test runner optimizations, most notably, preventing false sharing. The term \emph{false sharing} refers to a situation when independent variables get allocated on a single CPU cache line, which can easily prevent many weak behaviors. In our internal experiments, we have seen that using some method to avoid false sharing can significantly boost the frequency of weak behaviors, but unfortunately, we struggled to find a general solution to add into \litmuskt. We proposed adding a Kotlin/Native annotation similar to JVM's \code{@Contended}~\cite{youtrack-issue}. 

\section{Conclusion}

We built a litmus testing tool \litmuskt, which allows to write litmus tests in Kotlin and run them on different platforms. Using the techniques found in established litmus testing tools, the tool was able to discover new unsafe-publication-related bugs in the Kotlin/Native compiler. The Kotlin/Native team showed interest, and \litmuskt has been added into the Kotlin CI testing pipeline. We believe that our tool can be of use to both researchers and practitioners in the further study of the Kotlin's concurrency semantics.

\bibliographystyle{ACM-Reference-Format}
\balance
\bibliography{main}

\end{document}